# The True Cause of Magnetostriction

Yuri Mnyukh

*76 Peggy Lane, Farmington, CT, USA, e-mail: yuri@mnyukh.com*
(Updated: April 11, 2014)

___

**Abstract** The cause of magnetostriction is revealed by analyzing this phenomenon in a polydomain crystal of Fe. It is based on the two fundamentals: (a) magnetization is a rearrangement of spin directions by rearrangement of the crystal structure, and (b) the α-Fe has a tetragonal rather than cubic structure. The magnetostriction results from bringing the shorter tetragonal axis to coincide with, or closer to the direction of the applied magnetic field. It is not rooted in the alleged deformation of crystal unit cell.

___

## 1. Introduction

Magnetostriction was first observed by J. Joule in 1842. Currently it is typically described as "deformation of a body" as a result of its magnetization. The elongation or contraction in the direction of applied magnetic field $\pm (\ell-\ell_o) / \ell_o$ is usually between $10^{-5}$ and $10^{-3}$ and accompanied by the opposite sine changing in the transverse direction, so that the volume remains almost the same. The above definition is not incorrect, but we would prefer not to use the word "deformation", since it unwittingly suggests (erroneously as will be shown) to ascribe the phenomenon to deformation (also called "distortion") of its crystal lattice. We define the magnetostriction simply as change of shape of a ferromagnetic body by applied magnetic field.

Magnetostriction involves no change of crystal structure and not to be confused with the changes taking place in phase transitions, especially those driven by application of magnetic field. No credible theory of the phenomenon exists. Here is how the current status of the problem is described in [1]: "Despite the tremendous advances in modern electronic structure theory for studies in material science, magnetostriction has been rarely attacked until recently due to its intrinsic complexity". It will now be shown that the seeming complexity of the phenomenon results from looking for the answer in wrong direction. The secret of its origin is hidden in the *magnetization process*, provided it is properly understood.

The ample literature on magnetostriction is devoted to its observation in different materials, its technological applications, and its negative effects. Attempts of its explanation are rare, superficially-descriptive, brief and vague, but unanimous in the belief that magnetostriction results from elastic deformation / distortion of the crystal lattice strained by forces of magnetic interaction.

To illustrate these views we turn to two renowned authorities in ferromagnetism.

K. P. Belov [2]: *"Magnetization, which occurs by displacement of domains boundaries and rotation of domain magnetic moments, leads to changing of equilibrium distances between the atoms of the lattice; the atoms are shifted and magnetostrictive deformation of the lattice occurs"*.

This explanation is erroneous in every its point (see Sections 2 and 3). It incorrectly describes the mechanism of magnetization. It is not specified why interface propagation will change the inter-atomic distances (it will not). Rotation (reorientation) of magnetic moments in a domain is not possible. Deformation of the lattice was merely a guess: it does not occur.

S. Chikazumi [3]: *"The reason for* [magnetostriction] *is that the crystal lattice inside each domain is spontaneously deformed in the direction of domain magnetization and its strain axis rotates with the rotation of the domain magnetization, thus resulting in a deformation of the specimen as a whole"*.

Noticeably, here magnetization by motion of domain boundaries − an undeniable experimental fact − is not even mentioned. A further commenting on this excerpt would be a repetition.

In both cases (and everywhere else) the crystal lattice is assumed to become strained. The idea that magnetostriction is due to elastic deformation of crystal lattice is so entrenched that it even entered into this definition: [magnetostriction is] *"dependence of the state of strain (dimensions) of a ferromagnetic sample on the direction and extent of its magnetization"* [4]. No attention is being paid to the fact that removing the magnetic field does not lead to elastic relaxation of the sample towards its "unstrained" condition. Why? It is simply not strained.

## 2. Fundamentals of Ferromagnetism: New *vs.* Old

Magnetostriction cannot be correctly explained until the physics of magnetization is understood. The currently dominated theory of ferromagnetism, based on the Heisenberg's idea on extremely strong *electron exchange interaction*, has failed to account for thermodynamic stability of ferromagnetic state and for basic ferromagnetic phenomena [5,6]. As for the magnetization is concerned, the current theory involves a belief that it is realized by a "rotation" (meaning: reorientation) of spins *in* the crystal structure. The theory is powerless to explain why magnetization proceeds by propagation of domain boundaries.

The new physics of ferromagnetism was put forward in 2001 [5] and formulated again in [6]. Here are its main principles:

**1. Stability of ferromagnetic state.**
The Heisenberg's extremely strong "field of electron exchange forces" in a ferromagnetic materials does not exist. Only the classical magnetic interaction is real; its contribution to the crystal total free energy is relatively small as against the chemical crystal bonding (and by itself has usually a destabilizing effect). A simple consideration that the *total* free energy of a ferromagnetic crystal determines its stability eliminates the central problem the previous theory was unable to solve: a *ferromagnetic crystal is stable due to its low total free energy in spite of a possible destabilizing effect of the spin magnetic interaction.*

**2. Orientation of spins in their carriers.**
*Orientation of a spin is uniquely bound to the orientation of its atomic carrier.* The spatial distribution and orientation of spins (*i.e.* state of magnetization) is *imposed* by the crystal structure (by its chemical bonding). It also follows that any change of spin orientation in a crystal (in other words, any magnetization) requires reorientation of the carriers. That can occur only by restructuring of the crystal itself. The general mechanism of a crystal restructuring in solids is nucleation and molecule-by-molecule rearrangement at interfaces [5,7]. That is why any magnetization proceeds only by interface propagation, while spin "rotation" (reorientation) in the otherwise intact crystal lattice is impossible [5,6,8]. On the same reason all ferromagnetic phase transitions (i.e. those involving magnetization) are *structural*. They are "first order" rather than "second order" [5,9].

**3. Essence of paramagnetic state.**
Paramagnetic state is the orientation-disordered crystal (ODC) state known in other types of crystals, where a translation 3-D crystal order exists, but atoms or molecules − spin carriers − are engaged in a thermal rotation resulting in zero spontaneous magnetization.

## 3. Magnetization of Iron

The physical origin of magnetostriction will be demonstrated by showing how it emerges in iron. The phenomenon easily reveals itself in a common technical Fe. It is known that after melt solidification and cooling down to room temperature, the common solid Fe consists of arbitrarily shaped and oriented grains. Every grain is not a single crystal, but a *polycrystal* − a complex of peculiarly organized single-crystal domains separated by straight boundaries. It is a structural reorganization of the domain complex, and not a deformation of the Fe unit cell, will be shown to give rise to the magnetostriction.

Iron is a ferromagnetic material, but, as known, can be either in a ferromagnetic (F) or non-ferromagnetic (paramagnetic, P) state. Depending on the state, it reacts differently to magnetic field $H$. In the F-state the resultant magnetization ($M$) will be either $M \neq 0$, or changed from $M_1$ to $M_2$; in the P-state $M \equiv 0$ under all circumstances.

In terms of the principles listed in Section 2 we are able to demonstrate the structural mechanism of magnetization by using two neighboring domains by way of example (Fig. 1). They are single crystals of identical crystal structure, naturally magnetized to saturation, but differently oriented together with their magnetization vectors $M_S$. Application of magnetic field $H$ to this system cannot turn $M_S$ toward $H$ directly, considering that $M_S$ is fixed by crystal structure. It upsets the balance of the domain free energies $E$, initially identical. The domains react as if they are different crystal phases: the one of a lower free energy (No. 1) becomes preferable and grows by a movement of the interface AB to right into No. 2. *Magnetization of the system is realized without a "rotation" of the whole domains* (sometimes suggested, but unimaginable*) or reorientation of spins in the otherwise intact crystal structure* (which is impossible − as stated in Section 2).

The same approach accounts for $M \equiv 0$ when material is in its P-state. Application of magnetic field does not upset the energy balance $E_1 = E_2$ due to thermal rotation of *atoms* and their spins in both domains, so the interface remains still. However, if spins could freely change their orientations because not being bound to the orientations of their atomic carriers, at least some orientation effect (magnetization) toward $H$ due to direct magnetic interaction could be expected.

Concluding, it is to be noted that

(a) the described magnetization process is based on the same principle of crystal growth as that taking place in solid-state phase transitions, even though here no change in the type of crystal structure is involved, and

(b) the cause of magnetization by interface propagation gets its natural explanation.

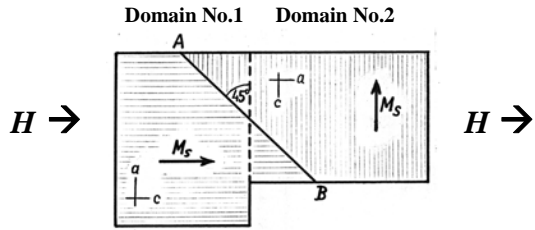

Fig. 1. Magnetization process in Fe by propagation of interface AB between the two spontaneously magnetized domains of identical crystal structure. The AB, called "90°- border" is a crystallographic twin of zero thickness where the structures perfectly match. Application of magnetic field *H* makes the free energy of No.1 lower than No. 2, causing it grow by AB movement to right, consuming No. 2. However, magnetization of a solitary crystal needs nucleation.

## 4. The Crystal Structure of Fe

The crystal structure of the room-temperature α-phase of Fe will be reexamined. Over the whole 20$^{th}$ century the phase behavior of Fe (Fig. 2, left side) looked enigmatic [10]. There were three different crystal phases α, γ and δ on the temperature scale. The room-temperature α-phase, believed to be truly cubic, bcc, *a* = 2.86645 Å at 20 °C [11], stretched up the temperature scale to 910 °C where it undergoes phase transition to γ-phase [12]. The α-phase above ~ 770 °C is paramagnetic, but ferromagnetic below it. The paramagnetic part of the α(bcc) phase is called "β-phase" only conditionally and could not be clearly categorized [13-16]: "Beta ferrite (β-Fe) and beta iron (β-iron) are obsolete terms for the paramagnetic form of ferrite (α-Fe)"…"Beta ferrite is crystallographically identical to alpha ferrite, except for magnetic domains"…"The beta 'phase' is not usually considered a distinct phase but merely the high-temperature end of the alpha phase field".

Others call the ferromagnetic α↔β change a "second order phase transition". It was in violation of what the second-order phase transitions assumed to be, because "transition from the ferromagnetic to the paramagnetic state is a phase transition of a second kind. At the Curie point, where the spontaneous magnetization disappears, the symmetry of a ferromagnet changes sharply" [17]. If so, can the ferromagnetic α(bcc) → α(bcc) transition be "second order"?

In terms of the conventional theory of ferromagnetism this picture was enigmatic indeed. It led Myers [10] to state: "Unusual and nontypical, elemental iron can provide the impetus for discussing … basic thermodynamic concepts and the phenomenon and theory of ferromagnetism".

The required clarification has come from the new concepts of ferromagnetism referred in Section 2. All phases on the temperature scale (Fig. 2), including the β-phase, are crystallographically different and transit to their neighbors by nucleation and rearrangements at interfaces. The α-phase occupies the temperature range only from ~769 °C down. All the phase transitions are first-order, exhibiting phase coexistence, hysteresis and latent heat − detected or not yet. The phases β, γ and δ are ODCs. They are paramagnetic due to thermal rotation of the atomic spin carriers.

The existence of the β → γ and γ → δ, phase transitions ODC → ODC above ~769 °C is an indication that the atomic spin carriers are not spheroid-shaped entities and their hindered rotation required more space with increasing temperature. On that background the CRYSTAL → ODC transition α → β occurring without changes in density, latent heat, and other characteristics of first-order phase transitions, if true, would be an anomaly. But the above characteristics of this transition were incorrect. As has been concluded in [5], the room-temperature α-phase emerges upon cooling from the β-phase by nucleation and growth of a new crystal structure. While this change was not noticed for a very long time, the latent heat − signature of a structural transition − was ultimately recorded [18].

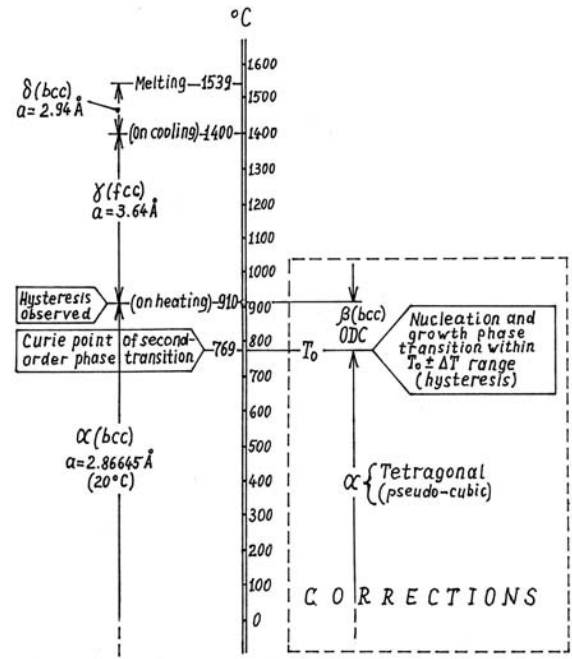

Fig. 2. Phase transitions in Fe as they believed to be. On the right - corrections eliminating existing contradictions. The difference is in the interpretation of the α-structure and α - β ferromagnetic transition at ~769 °C. ODC is paramagnetic orientation-disordered crystal state.

We shall take a close look at the crystal lattice of α-Fe within a domain. Its unit cell and the directions of spontaneous magnetization are shown in Fig. 3. *The spin magnetic interaction is an integral part of the forces establishing equilibrium interatomic distances. The structure is not strained. It is in its natural stable state.* While its symmetry was deemed to be cubic on the basis of precision measurements of the unit cell parameters, it cannot be so according to the rules of symmetry. Indeed, the uniform orientation of spins results from the uniform orientation of their atomic carriers located in the sites of the unit cell. Such crystal lattice lacks of cubic centre of symmetry, leaving it to be a pseudo-cubic. The unit cell is not truly cubic geometrically as well. The mutual repulsion of the parallel elementary dipoles in the two equivalent orthogonal directions $a$ and $b$ of the pseudo-cubic unit cell makes its $c$-parameter in the third orthogonal direction the shortest one, thus producing a tetragonal unit cell with $c < a = b$. The difference is minor, but it is sufficient to produce the observed magnetostriction.

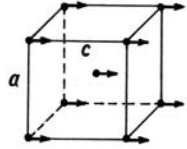

Fig. 3. Unit cell of ferromagnetic α-Fe crystal and its spontaneous magnetization. It is tetragonal with $c < a = b$. The difference is very small, but enough to be the source of the observed magnetostriction.

## 5. The Domain Structure of Fe

Every grain of technical Fe is a polydomain crystal comprised of single-crystal domains arranged in a specific pattern of a kind shown in Fig. 4. They are divided by straight twin interfaces. Applied magnetic field causes the interfaces to move in coordination. In three dimensions, the spins of every domain belong to one of six mutually orthogonal directions (Fig 5). At that, the spins of two neighboring domains form either 90° ("90° boundary") or 180° ("180° boundary").

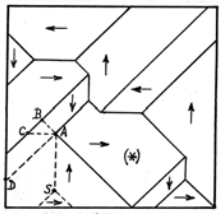

Fig.4 A schematic illustrating the features of domain patterns in the Fe polydomain crystals (shown is (*100*) plane). The broken lines help to comprehend their "rules": only interface AS is correct ([5], Sec. 4.9).

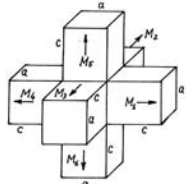

Fig. 5. Six mutual directions of spontaneous magnetization $M_S$ that make 90° and 180° to one another in a polydomain iron crystal. While the unit cell dimensions $c$ and $a = b$ are very close, it is axis $c$ which represents the $M_S$ direction. (The figure is not intended to illustrate the orientations of domain boundaries).

In a polydomain Fe crystal the $c$-axes of the domains are initially distributed over the six equivalent spatial directions. In the applied magnetic field the domain structure undergoes rearrangements by growth of the "better oriented" domains at the expense of others. In average, the domain structure acquires preference of its $c$-axis in the direction closer to the applied field ***H***, somewhere between 0° and 45° to ***H***, depending on the $H$ strength and other conditions.

## 6. Magnetostriction is Illustrated

Misinterpretation of magnetization, as described above, has been a major barrier to the identification of the cause of magnetostriction. Another obstacle, now also removed, was not detecting the unit cell of α-Fe to be a tetragonal rather than cubic.

The cause of the magnetostriction in a Fe polydomain crystal is illustrated in Fig. 6. Bozorth [19] indicated that a "magnetization reversal" does not produce magnetostriction, but changing of the spin direction by 90° does. The part A,B of Fig. 6 explains that. The $c$-axes of the domains in the case of 180° boundary remain aligned along ***H*** both before and after the crystal rearrangement. No geometrical change is produced. But the structural rearrangement resulting in alignment of the $c$-axes along ***H*** in case of the 90° boundary gives rise to changing of the body length.

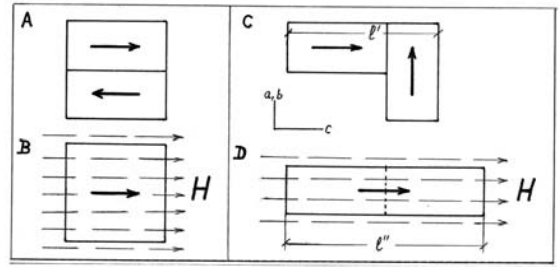

Fig. 6. Sketch showing a system of two domains of the shape reflecting their tetragonal unit cell. For the sake of illustration the parameter $c$ is made much longer than $a = b$. (A,B): Magnetostriction of the antiparallel domains is zero, as actually found experimentally. (C,D): In case of the domains making initially 90° with each other, rebuilding the domain on the right into a position with $c \parallel H$ gives rise to the magnetostriction $\ell'' - \ell'$.

## 7. Why Tetragonal Fe Has Not Been Detected

Quantitatively, difference between the $c$ and $a$ parameters was not large enough to be directly detected. But it was sufficient to be seen and measured as a cumulative effect called magnetostriction. Inasmuch as this phenomenon is a change in the shape of the same quantity of the matter, the length magnetostriction is accompanied by a transverse magnetostriction of the opposite sign, leaving very little to the *volume* magnetostriction. The latter is, evidently, a secondary effect accompanying any structural rearrangement.

We can roughly estimate the difference between the *c* and *a* parameters of an α-Fe unit cell by ascribing it to its reorientation by 90º. The maximum length magnetostriction (defined as fractional elongation or contraction) in Fe, reported in a number of experimental studies, including performed on single crystals, was about $2 \cdot 10^{-5}$, which gives us

$(a - c) = 2.86645$ Å $\cdot 2 \cdot 10^{-5} \approx 0.00006$ Å.

The value $a = 2.86645$ Å of a Fe unit cell, supposed to be cubic, was the result of special precision measurements in 1948 [11] by the best technique available at that time. It was a powder X-ray photography by recording back reflections on film in circular camera. The reported probable error of a single observation was ±0.00003Å with Co radiation and ±0.00006 Å with Cr radiation. Thus our estimate for $(a - c) \approx 0.00006$ Å, even if it is several times greater in reality, counts at the very limits of the precision given for $a = 2.86645$ Å. More importantly, the reflections were indexed in a cubic lattice, so an excessive widening of the reflection line (and those illustrated in the work are far ftom being sharp) was not taken into consideration. At that, the center of this line, being, evidently, a superposition of two partially overlappind lines, could be measured with high precision. A a result, the deviation of the a-Fe unit cell from a geometrically cubic could easily be overlooked.

## 8. "Electrostriction" is Differently Defined

The counterpart of magnetostriction in ferroelectrics had to be "*electrostriction*", but this is not the case. The latter notion was defined differently − not specific to ferroelectrics. As opposed to the definition of magnetostriction, this name includes a small elastic deformation of any dielectrics when they are placed in an electric field *E*. Because the (+) and (-) charges of the electric dipoles are, or can be, spatially separated in the crystal unit cell of a dielectric, application of *E* either causes its polarization by crystal rearrangement, or affects the value of its dipole moments. The latter produces some elastic deformation of the crystal. The analogous effect - induced magnetization and the corresponding elastic deformation - is essentially absent in ferromagnetic where the poles of the elementary magnetic moments cannot be split apart.

Even though the theory does not have a name for the ferroelectric counterpart of magnetostriction, it is real. Let us name it *ferroelectrostriction* (FES). The essence of what was said about magnetostriction holds true for FES. The case of $BaTiO_3$, for instance, is rather similar to that of Fe. The cubic paraelectric phase of $BaTiO_3$ becomes tetragonal below 120 ºC after its (nucleation and growth) phase transition into the ferroelectric phase. The only difference is that the *c*-parameter differs from the $a = b$ by 1%, *i.e.*, much greater than in Fe. Formation of the domain structure by nucleation in the six fixed orientations within the cubic paraelectric matrix occurs in the same manner. Domain rearrangements during polarization in electric field give rise to FES. The FES is a change in the shape of a polydomain ferroelectric body as a result of crystal rearrangements of the domains in the process of polarization under the action of applied electric field. In the combined effect of the induced polarization and FES, the latter is greater.

## 9. Conclusion

The simple crystal structure of Fe and the simple pattern of its spontaneous magnetization were very helpful in analyzing and illustrating the cause of magnetostriction.

The previous views on this phenomenon as deformation / distortion of crystal structure by magnetic field were based on misinterpretation of a magnetization process as reorientation of spins in the crystal lattice. On this reason a part of the present article was devoted to that important aspect of ferromagnetism. It was illustrated that magnetization is actually realized by a domain reconstruction at their interfaces according to nucleation-growth principle.

Another critical component in our explanation of the magnetostriction was to reveal a tetragonal rather than cubic crystal structure of Fe. It is the restructuring of the domain system toward predominant alignment of the tetragonal axes *c* in the direction of the applied magnetic field that gives rise to the change of the body dimension. No intrinsic deformation / distortion of crystal lattice involved in that process.

## REFERENCES

[1] R. Wu, V.I. Gavrilenko and A.J. Freeman, in *Modern Trends in Magnetostriction Study and Applications,* Ed. M.R.J. Gibbs, Kluwer Acad. Press (2000).

[2] K. P. Belov, *Magnetostriction,* in *The Great Soviet Encyclopedia, 3rd Edition (1979).*

[3] S. Chikazumi, *Physics of Ferromagnetism*, Oxford Univ. Press (2009).

[4] *McGraw-Hill Dictionary of Scientific & Technical Terms*, 6th Ed. (2003).

[5] Y. Mnyukh, *Fundamentals of Solid-State Phase Transitions, Ferromagnetism and Ferroelectricity*, 2001 [or 2$^{nd}$ (2010) Edition].

[6] Y. Mnyukh, *Amer. J. Cond. Mat. Phys*. 2012, 2(5): 109.

[7] Y. Mnyukh, *Amer. J. Cond. Mat. Phys*. 2013, 3(4): 89.

[8] Y. Mnyukh, arxiv.org/abs/1101.1249.

[9] Y. Mnyukh, *Amer. J. Cond. Mat. Phys*. 2013, 3(2): 25.

[10] C. E. Myers, *J. Chem. Educ.* 43, 303–306 (1966).

[11] D E Thomas, *J. Sci. Instrum.* **25** (1948) 440.

[12] J. Donohue, *The Structure of Elements*, John Wiley & Sons (1974).

[13] D. K. Bullens *et al*., *Steel and Its Heat Treatment*, Vol. I, 4$^{th}$ Ed., J. Wiley & Sons Inc., 1938.

[14] S. H. Avner, *Introduction to Physical Metallurgy,* 2$^{nd}$ Ed., McGraw-Hill, 1974.

[15] *ASM Handbook, Vol. 3: Alloy Phase Diagrams*, ASM International, 1992.

[16] B. D. Cullity & C. D. Graham, *Introduction to Magnetic Materials,* 2$^{nd}$ Ed.*,* IEEE Inc.

[17] S.V. Vonsovskii, *Magnetism*, v. 2, Wiley (1974).

[18] Sen Yang *et al.*, *Phys. Rev.* B 78 (2008) 174427.

[19] R.M. Bozorth, *Ferromagnetism*, Van Nostrand (1951).